# A 12-Step Process for Industrial Internet of Things (IIoT) Forensics

Victor Kebande, *University of Colorado, Boulder, USA,*
*University of Colorado, Denver, USA,*
*Blekinge Institute of Technology, Sweden*

*Abstract—The increasing deployment of the Industrial Internet of Things (IIoT) in critical infrastructure sectors like manufacturing, healthcare, and transportation has shown new challenges for Digital Forensics (DF). Traditional DF methodologies are not well equipped to handle the complexity, scale, and heterogeneity of IIoT environments. This paper introduces a comprehensive Twelve-Step Process (TSP) tailored specifically for IIoT incidents, addressing the need for effective investigation and Potential Digital Evidence (PDE) handling in such dynamic environments in DF. We begin by exploring the importance of IIoT and its role in industrial ecosystems, followed by an examination of existing DF challenges. Each step of the process, from forensic readiness to investigation closure, is designed to ensure robust PDE collection, analysis, and legal compliance to increase chances of admissibility from a DF scenario.*

## introduction

The fourth industrial revolution, industry 4.0 has in recent past propelled the Industrial Internet of Things (IIoT) which has also revolutionized industries. IIoT has enabled a seamless connectivity between machines, devices, and systems, facilitating greater efficiency, automation, and data where data has played a central role. From smart factories to autonomous vehicles, IIoT systems are deeply embedded in critical infrastructure, making their security and integrity paramount. However, as these systems become more integrated and complex, they also present significant challenges in terms of cyber threats, data breaches, and system vulnerabilities. When incidents occur, the ability to investigate and recover digital evidence in an accurate, systematic, and legally compliant manner is crucial for mitigating risks and ensuring accountability.

Digital Forensics (DF), which is a field that traditionally is concerned with the investigation of cybercrimes in the context of personal computers and network systems, has hardly been integrated and has struggled to keep pace with the evolving landscape of IIoT. In fact, the existing DF processes, which are designed for conventional Information Technology (IT) and Operational Technology (OT) environments have been seen to often fail to address the unique challenges posed by IIoT systems. These challenges range from device heterogeneity, real-time data generation, to complex distributed architectures. As such, this has led to a gap in effective forensic processes that tailored specifcally to the IIoT domain.

The traditional digital forensics also faces well-known challenges such as evidence volatility, device diversity, chain-of-custody complexity, and scalability limits, which have become even more prevalent in IIoT environments and this motivates the need for a specialized forensic process.

The goal of this paper, therefore, is to introduce a structured, twelve-step process for IIoT forensics, that has been designed to provide DF researchers and professionals with a generic comprehensive framework for handling potential security-related incidents across IIoT environments. In the context of this paper, IIoT forensic process refers to a structured and systematic set of prescribed procedures designed to collect, preserve, analyze, and present digital evidence from IIoT systems. By examining existing digital forensic frameworks in IIoT ecosystems, identifying gaps, and building upon established best practices, this paper aims to offer a practical guide for investigators on how to efficiently collect, preserve, and analyze digital evidence from IIoT devices and networks.

The contributions of this paper are as follows:

- Propose a twelve-step approach to IIoT Forensics
- Provide a contextual evaluation of the study and show the strengths and weaknesses of existing IIoT Forensic frameworks.

The remainder of the paper is organized as follows: In Section II, a brief Background and Motivation of exploring IIoT is given. This is followed by Related Work Section III. Next, the Methodology follows in Section IV. This is then followed by the Twelve-step processes for IIoT Forensics in Section V, with key discussions in Section VI. The paper concludes and mentions a future work in Section VII.

## Background and Motivation

### IIoT: Importance and Role

The Industrial Internet of Things (IIoT) is seen as a critical component of Industry 4.0, given that traditional industrial systems are augmented with intelligent, networked devices [1]. These devices, include sensors, machines, and control systems, and are able to communicate in real time, creating interconnected ecosystems that allow for optimized decision-making, predictive maintenance, and resource management. Also, IIoT applications span a wide range of industries like manufacturing, energy, healthcare, transportation, and agriculture [2]. IIoT differs from the conventional IoT by operating within industrial, safety-critical, and cyber-physical environments where connected sensors, controllers, and machines support real-time operational processes, making their forensic requirements uniquely demanding.

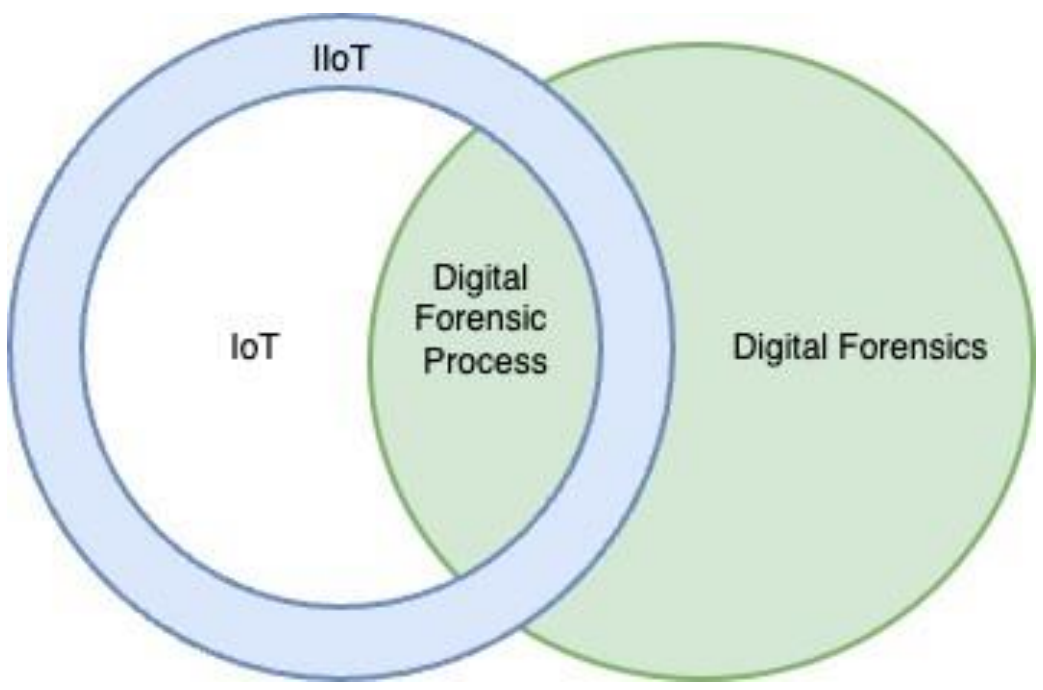

**FIGURE 1.** Interdependencies and Relationship between IoT, IIoT, Digital Forensics and Digital Forensics Processes

The role that IIoT plays is important, given that it modernizes industrial operations, thus improving operational efficiencies, reducing downtime, and enabling the collection of vast amounts of data that provide valuable insights into system performance.

However, this large expansion of IIoT and the interconnected devices also introduces new challenges, particularly in cybersecurity, privacy, and data integrity. It has been seen that IIoT and OT devices have become more critical to the functioning of industries. With the changing cyber threat landscape, the need for robust digital forensic methods to investigate incidents and breaches involving these devices becomes increasingly important.

Figure 1 shows the interdependencies and relationships among the IoT, IIoT, DF, and DF processes. The diagram highlights how IIoT is able to extend the scope of IoT. This can be achieved by integrating industrial systems, which overlap with DF processes due to the increasing need for security and DF and attribution in interconnected environments.

### Need for IIoT Forensic Processes

The increasing integration of IIoT with digital forensics has been seen to create new avenues for enhancing security and evidence handling in industrial ecosystems, which can only be realized with prescribed and acceptable DF processes. However, this integration has led to a rise in vulnerabilities, which highlights the urgent need for specialized DF processes tailored to IIoT environments. Traditional forensic methodologies have often failed to address the complexity, heterogeneity, and real-time demands of IIoT systems.

This has necessitated the creation of a more robust and structured approach. IIoT ecosystems, characterized by interconnected devices and real-time operations, are increasingly prone to vulnerabilities that traditional forensic frameworks cannot adequately address [13]. This integration of IIoT and digital forensics is essential to bridge this gap, ensuring the sanctity, admissibility, and effective handling of digital evidence. Traditional DF approaches rely on stable endpoints and centralized evidence, which contrasts sharply with IIoT's real-time, heterogeneous, and distributed forensic landscape, making classical methods insufficient without adaptation.

Motivated by these shortcomings, this study is positioned to introduce a structured IIoT forensic process that emphasizes three distinct elements: process models to guide investigations, need for standards and frameworks to ensure uniformity and compliance, and enhanced investigative techniques to improve analysis depth and scope. In addition, it is important to have robust incident assessment and analysis to enable effective real-time responses. In the long run, it is expected that this study will provide a comprehensive solution to safeguard IIoT systems and lays the foundation for advancing research and practice in IIoT forensics.

**TABLE 1.** Challenges in IIoT Forensics

| Challenge | Description | Impact on IIoT Forensics | REF |
|---|---|---|---|
| **Data Volume and Complexity** | IIoT environments generate vast amounts of diverse data (e.g., sensor data, logs, system data). | The sheer volume and variety of data make it difficult to analyze and extract meaningful evidence | [3] |
| **Device Heterogeneity** | IIoT networks consist of a wide range of devices, from legacy systems to modern IoT devices. | Forensic methods must be adaptable to different device types, requiring a diverse set of tools and knowledge | [4] |
| **Real-time Data Processing** | Many IIoT systems require real-time data processing to ensure operational continuity. | The need for immediate data analysis and response complicates forensic investigations, as evidence may be overwritten or lost | [5] |
| **Data Integrity and Authenticity** | IIoT data is often transmitted over multiple networks and processed by various systems, which may impact data integrity. | Ensuring that the collected data has not been tampered with or altered is essential for maintaining the validity of the evidence | [6] |
| **Encryption and Privacy** | IIoT systems often use encryption for secure communication. | Forensic investigators may face difficulties accessing encrypted data without violating privacy regulations or security protocols | [7] |
| **Lack of Standardized Forensic Processes** | There are no universally accepted forensic processes tailored to IIoT environments. | Investigators may lack proper guidelines, leading to inconsistent or incomplete forensic investigations | [8] |
| **Legal and Regulatory Challenges** | Different regions have varying laws and regulations regarding data privacy, security, and evidence handling. | Navigating these legalities can complicate the collection, analysis, and presentation of forensic evidence | [9] |
| **Network Complexity and Security** | IIoT systems are often interconnected across multiple networks, which may introduce vulnerabilities. | Investigating network-related incidents requires a deep understanding of the network architecture and potential attack vectors | [10] |
| **Lack of Skilled Forensic Professionals** | The specialized nature of IIoT forensics requires professionals with expertise in both digital forensics and industrial systems. | There is a shortage of qualified professionals, which can delay or hinder forensic investigations in IIoT environments | [11] |
| **Device Autonomy and AI Integration** | Many IIoT devices use AI algorithms to make autonomous decisions. | Investigating incidents involving AI-driven devices can be challenging, especially when the decision-making process is opaque or difficult to trace | [12] |

## Digital Forensic Challenges in IIoT

IIoT introduces a range of forensic challenges due to its unique characteristics, which have also been summarised in Table 1. These challenges include the massive scale, heterogeneity, and complexity of devices and networks. In terms of device and network heterogeneity, IIoT deployments typically integrate hundreds to thousands of heterogeneous endpoints in a single industrial site. These include, for example, industrial sensors, programmable logic controllers (PLCs), remote terminal units (RTUs), smart meters, autonomous mobile robots, and industrial gateways sourced from multiple vendors within industrial control system (ICS) environments. [14]. Many of these components rely on proprietary or vendor-specific communication protocols, which complicates consistent data collection and interpretation for forensic purposes, especially in industrial control system (ICS) environments. At the same time, the high rate of telemetry, log, and control data generated in real time produces large data volumes that can easily overwhelm traditional forensic tools, reinforcing the need for specialized techniques for data extraction and analysis.

In a time-sensitive environment, the rapid acquisition of data is essential to prevent crucial evidence from being overwritten or lost, adding to the urgency of forensic investigations. The limited forensic support in IIoT devices is another key challenge, as many devices do not have built-in logging capabilities or secure storage, making it difficult to capture and preserve evidence. Moreover, IIoT systems are often integrated into a broader cyber-physical system (CPS), where physical elements, like industrial machinery, which interact with digital systems, further complicating the forensic process. It is worth noting also that, distributed and decentralized systems in IIoT environments mean that evidence may be fragmented across multiple devices, networks, and cloud environments, requiring advanced correlation and looking or raiding the artifacts using some techniques [15]. Another pressing challenge is the lack of standardized forensic procedures, as most forensic tools and methodologies are not designed for IIoT environments [8]. Also, the legal and ethical considerations come into play, especially when dealing with cross-border, multiple jurisdictions, data and privacy concerns, as IIoT systems often span multiple jurisdictions with differing laws regarding data access and preservation.

## Related Work

The need for IIoT forensic process that been identified as a forgotten mantra that need to be addressed by [16]. In addition, an industrial analysis framework for IIoT has been suggested by [17] that is able to enumerate and characterize IIoT devices based on the system architecture in order to analyze security threats and vulnerabilities. While this study gives practical classification schema, digital forensics aspect has hardly been a focus. Also, a blockchain has been incorported as a measure for addressing digital evidence collection in IIoT concerns by [18] using Delegated Proof of Stake (DPoS). In this study, it has been observed that blockchain's application is relevant for judicial access purposes and digital evidence storage, however, a process model has hardly been suggested to back this up. Consequently, a cross-layer forensic investigation approach has been suggested for IIoT that has a focus on device attacks in critical infrastructure applications. In this study, a forensic framework for IIoT that had pillars for higher layer digital and lower layer physical (PHY) forensic was suggested [11]. While this study is very informative, this framework does not offer strategies for standardized processes. In spite of that, research that has a focus on the need for securing IIoT has focused on DFR framework and cybersecurity approach, and the authors have identified challenges and further developed a framework for IIoT networks which is aligned to an attack model. It has been observed that the proposed DFR framework has adopted a seven-step process. while this seven-step process is relevant, it's focus is mainly on pre-incident processes and does not incorporate post-incident processes [19]. In addition to this, jurisdictional and digital forensic challenges for IIoT by [20] where the need for formalizing an approach for IIoT Standardization and protocols has been mentioned as a prerequisite. Other studies on the need for standardizing IIoT forensic process has been suggested by [8]. Also, a study by [13] has identified, Lack of certified forensic incident response tool and methodologies in IIoT ecosystems as a crucial step when assessing the security and digital forensics for IIoT environments...

## The Twelve-Step Process for IIoT Forensics

This section provides a detailed account of the proposed twelve-step process designed for IIoT forensics as a contribution. Each of the mentioned step in this approach is crucial for the success of the forensic investigation and addresses the unique characteristics and challenges of IIoT systems as is shown in Table 1. The suggested twelve-step process for IIoT forensics is an iterative process model that consist of the following steps that have been shown in Figure 4: Forensic Readiness (1), Initial Incident Assessment (2), Evidence Collection and Preservation (3), IIoT Device and Network Identification (4), Data Acquisition from IIoT Devices (5), Forensic Imaging and Integrity Checks (6), Analysis of Communication Protocols and Traffic (7), Anomaly Identification (8), Data Correlation (9), Legal and Ethical Consideration in Evidence Handling (10), Forensic reporting and Documentation (11) and Investigation Closure (12). Each of the above-mentioned step has been discussed next.

1) **Step 1: Forensic Readiness**
   *Forensic readiness* involves preparing an organization's infrastructure, systems, and personnel to respond to potential incidents before they occur and for IoT ecosystems respectively. This step shown in Figure 2 includes setting up appropriate logging mechanisms, ensuring that devices are configured in a way that allows the collection of relevant evidence, and establishing protocols for quickly identifying and addressing security breaches. An organization's readiness to investigate IIoT-related incidents can significantly reduce the time spent gathering evidence and can improve the overall effectiveness of forensic analysis. This is factored as a pre-incident planning and preparation before Incident identification as highlighted by ISO/IEC 27043: International standard.
2) **Step 2: Initial Incident Assessment**
   The *Initial Incident Assessment* is a critical first step in any forensic investigation. Also, as a standard step in ISO/IEC 27043, it involves determining the scope and severity of the incident, identifying the devices and systems affected, and establishing an initial response plan. This phase includes gathering preliminary information about the event, such as the time of occurrence, the nature of the attack (if known), and the impact on operations. Early identification of key assets and critical devices within the IIoT infrastructure helps forensic experts to prioritize their actions and allocate resources appropriately. Preliminary steps include initilization, acquisitive and investigative processes.
3) **Step 3: Evidence Collection and Preservation**
   *Evidence Collection and Preservation* are fundamental in IIoT forensics to ensure that all relevant data is gathered accurately and securely. This step involves identifying and capturing data from various sources, including IIoT devices, gateways, communication networks, and cloud services. It is vital that evidence is preserved in a manner that maintains its integrity and admissibility in court. Forensic investigators must ensure that data is collected without altering the original state of the systems and devices involved. In IIoT environments, this may involve capturing logs and telemetry from programmable logic controllers (PLCs), industrial sensors, remote terminal units (RTUs), industrial gateways, and supervisory control and data acquisition (SCADA) systems that store operational data relevant to forensic investigations.

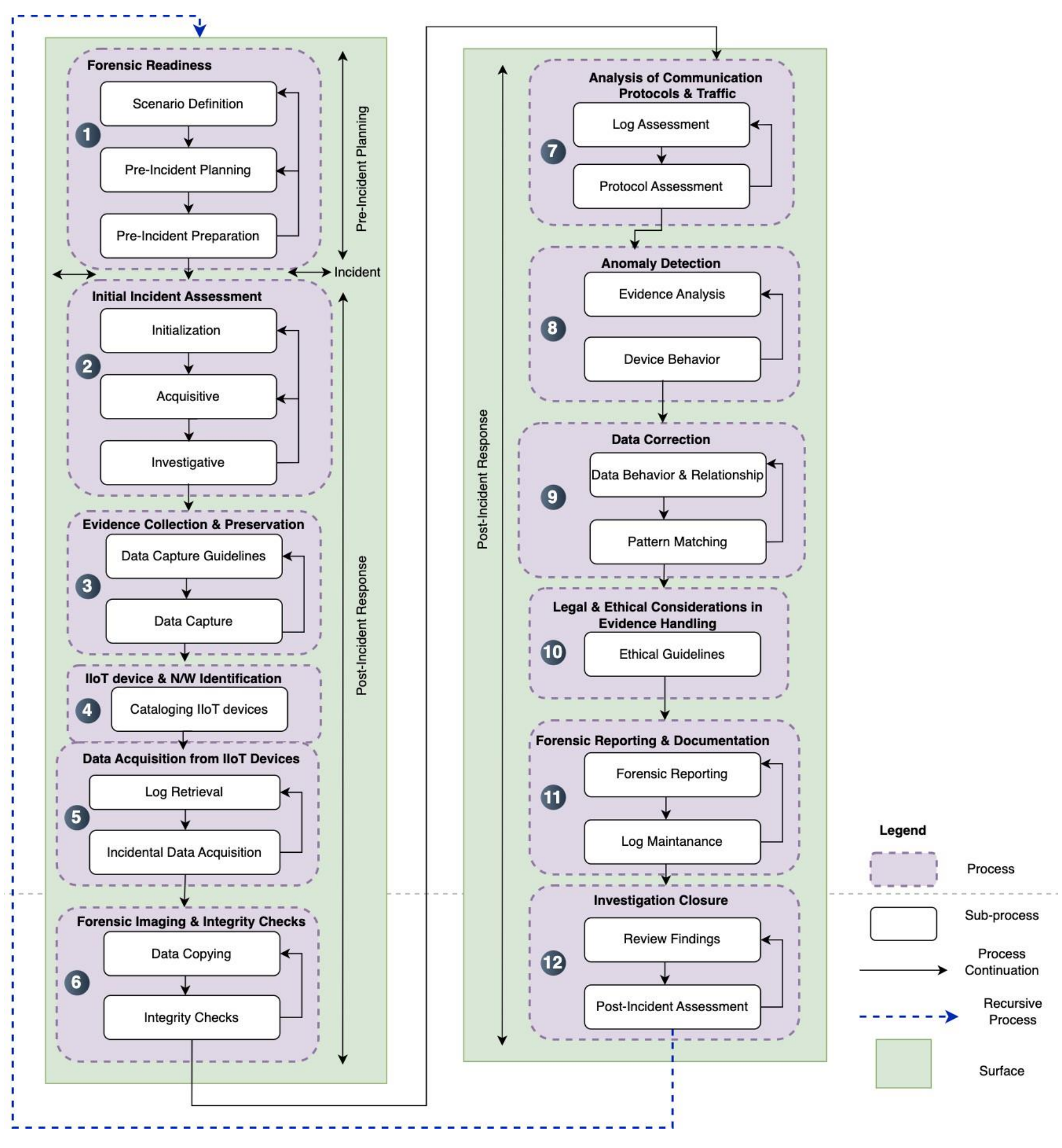

**FIGURE 2.** A Twelve-Step Process for Industrial Internet of Things (IIoT) Forensics

4) **Step 4: Identification of Devices and Networks Involved**
*Identifying the devices and networks* involved in an incident is crucial for understanding the full scope of the event. This process includes cataloging all IIoT devices (e.g., sensors, actuators, controllers) and networks (e.g., wireless, wired, mesh networks) that may have been affected or compromised. Mapping out the infrastructure allows forensic investigators to determine how the attack propagated through the system, which devices were impacted, and any potential vulnerabilities exploited during the incident.

5) **Step 5: Data Acquisition from IIoT Devices**
*Data acquisition* from IIoT devices involves extracting relevant data from the devices involved in the incident. This step typically includes retrieving logs, sensor readings, configuration files, and any other stored data that can provide insight into the incident. Forensic investigators need to use specialized tools and methods to access data from a variety of IIoT devices, including those that may be embedded or have limited interfaces for external access. Examples include retrieving telemetry logs from PLCs, sensor readings from industrial monitoring systems, configuration data from industrial gateways, and device state information exposed through industrial communication protocols such as Modbus, OPC-UA, or MQTT.

6) **Step 6: Forensic Imaging and Data Integrity Checks**
   *Forensic imaging* involves creating exact copies (or images) of the data from the affected devices, ensuring that the original data remains unaltered. This step is essential for maintaining the *integrity of the evidence* and enabling further analysis. Data integrity checks, such as hashing, are performed on these images to verify that no tampering or corruption has occurred during the acquisition process. This ensures that the collected data can be used as credible evidence in legal or regulatory proceedings.
7) **Step 7: Analysis of Communication Protocols and Traffic**
   *Analysis of communication protocols and traffic* provides valuable insights into how data was exchanged between devices and systems during the incident. In IIoT environments, this often involves examining industrial communication protocols such as Modbus, OPC-UA, DNP3, or MQTT that are commonly used between sensors, PLCs, industrial gateways, and SCADA systems.
8) **Step 8: Identification of Malicious Activity or Anomalies**
   In this step, forensic investigators analyze the collected evidence to identify signs of malicious activity or anomalies, such as unauthorized access, data manipulation, or malware presence. In industrial environments this may include detecting abnormal sensor telemetry patterns, unauthorized modifications of PLC logic, or unusual communication patterns between IIoT gateways and cloud-based monitoring platforms.
9) **Step 9: Correlation of Data Across Devices and Systems**
   *Data correlation* across devices and systems helps forensic investigators identify relationships between different pieces of evidence and understand the broader impact of the incident. For example, investigators may correlate PLC logs, sensor telemetry, gateway communication records, and SCADA event logs to reconstruct the sequence of events during an industrial incident.
10) **Step 10: Legal and Ethical Considerations in Evidence Handling**
   Handling evidence in a *Legal and Ethical* manner is essential for ensuring that the findings from the forensic investigation are admissible in court. This step involves adhering to legal frameworks, such as data privacy laws, and ensuring that evidence is managed according to established protocols. Investigators must also respect the privacy rights of individuals and organizations, ensuring that their actions do not violate any laws or ethical guidelines during the investigation.
11) **Step 11: Reporting and Documentation**
   *Reporting and Documentation* are crucial for maintaining transparency and accountability throughout the forensic process, especially standardized reports. This step involves compiling detailed reports that document the findings of the investigation, the methods used, and the conclusions drawn. It is essential that reports are clear, concise, and accessible for various stakeholders, including legal teams, organizational leadership, and regulatory authorities. Documentation also includes maintaining logs of all actions taken during the investigation to ensure a complete audit trail.
12) **Step 12: Investigation Closure**
   The *Investigation Closure* step shown in Figure 2 marks the end of the forensic investigation. It involves reviewing the findings, finalizing reports, and ensuring that all evidence has been properly handled and stored. The closure process also includes a post-incident review to assess the effectiveness of the forensic investigation and identify areas for improvement in future cases. Additionally, it may involve providing recommendations for mitigating the risks of future incidents and improving the overall security posture of the IIoT environment.

## Discussions

The twelve-step process for IIoT forensics provides a structured and comprehensive approach to handling forensic investigations in IIoT environment. In addition, the twelve-step process for IIoT forensics is an iterative process model that consist of the following steps that have been shown in Figure 2 as follows: Forensic

Readiness (1), Initial Incident Assessment (2), Evidence Collection and Preservation (3), IIoT Device and Network Identifation (4), Data Acquisition from IIoT Devices (5), Forensic Imaging and Integrity Checks (6), Analysis of Communicationn Protocols and Traffic (7), Anomaly Identifation (8), Data Correlation (9), Legal and Ethical Consideration in Evidence Handling (10), Forensic reporting and Documentation (11) and Investigation Closure (12).

The evaluation of these processes highlights the potential to streamline investigations by offering clear guidelines for conducting digital forensic investigations in IIoT environments. In particular, the processes addresses key challenges unique to IIoT environments, such as device heterogeneity, data integrity, and the dynamic nature of IIoT systems as it was shown in previously in Table 1. Consequently, these processes also improve upon existing frameworks by emphasizing forensic readiness which is a crucial pre-incident and evidence preparation phase, legal and ethical considerations, and cross-device data correlation, which is a suitable approach in the complex IIoT environments. Consequently, it is expected that the inclusion of forensic readiness is projected to maximize the use of digital forensic investigation, while minimizing the cost needed to conduct a reactive (investigation) process. It is worth noting that this process translated and fits to some of the prescribed processes that have been mentioned in ISO/IEC 27043 International standard, which defines approaches for Information technology — Security techniques — Incident investigation principles and processes.

This twelve-step process approach acts as blueprint for conducting extensive investigations across IIoT ecosystems while minimizing disruption to ongoing industrial operations. The practical implications of adopting the above-mentioned twelve-step IIoT Forensic approaches include the need for specialized forensic tools and expertise to handle the complex, real-time data generated by IIoT devices. While the twelve-step process is comprehensive, challenges remain in its implementation, particularly in large-scale IIoT systems, where issues such as scalability, integration with legacy systems, and ensuring data integrity in real-time pose significant obstacles.

## Conclusion and Future Work

This paper proposed a twelve-step process for IIoT forensics, designed to address these challenges by providing a structured and systematic approach to forensic investigations. The twelve-step process emphasizes essential aspects such as forensic readiness, evidence collection, data integrity checks, analysis of communication protocols, and legal considerations. By ensuring the thorough investigation of incidents in IIoT environments, this process can help forensic professionals maintain the integrity of the evidence and ensure that investigations are effective, efficient, and legally compliant.

Future work aims to implement that 12-step approaches in practical use-case, in order to provide the proof of this concept that has been discussed throughout this paper.

**VICTOR KEBANDE** is a cybersecurity researcher and assistant professor of ITsecurity in secure distributed systems at the Blekinge Institute of Technology,371 79 Karlskrona, Sweden; ATLAS Institute, University of Colorado Boulder, Boulder, CO, USA; and University of Colorado Denver, Denver, CO, USA. His research interests include cybersecurity, digital forensics in the Internet of Things, artificial intelligence in cybersecurity, critical infrastructure protection,and cloud security. Kebande received a Ph.D. in computer science (informationand computer security architectures and digital forensics) from the Universityof Pretoria. He serves on the Editor of Forensic Science International: Reports. He is a Member of IEEE. Contact him at victor.kebande@bth.se.